\documentclass[aps,prd,preprintnumbers,showpacs]{revtex4}
\usepackage[dvips]{graphicx}
\begin{document}

%
%

\eprint{Nisho-3-2014}
\title{Axion Stars and Fast Radio Bursts}
\author{Aiichi Iwazaki}
\affiliation{International Economics and Politics, Nishogakusha University,\\ 
6-16 3-bantyo Chiyoda Tokyo 102-8336, Japan.}   
\date{Oct. 16, 2014}
\begin{abstract}
We show that fast radio bursts arise from collisions between axion stars and neutron
stars. The bursts are emitted in the atmosphere of the neutron stars.
The observed frequencies of the bursts are given by the axion mass $m_a$ such as
$m_a/2\pi\simeq 1.4\,\mbox{GHz}\,\big(m_a/(6\times 10^{-6}\mbox{eV})\big)$.
From the event rate $\sim 10^{-3}$ per year in a galaxy, 
we can determine the mass $\sim 10^{-11}M_{\odot}$ of the axion stars. 
Using these values we can explain short durations ( $\sim $ms ) 
and amount of radiation energies ( $\sim 10^{43}$GeV ) of the bursts.
\end{abstract}
\hspace*{0.3cm}

\hspace*{1cm}

\maketitle

Fast Radio Bursts ( FRBs ) have recently been discovered\cite{frb,frb1,frb2} at around $1.4$ GHz frequency.
The durations of the bursts are typically
a few milliseconds. The origin of the bursts has been suggested to be
extra-galactic owing to their large dispersion measures.
This suggests that the large amount of the energies $\sim 10^{46}$GeV/s
is produced at the radio frequencies.
The event rate of the burst is estimated to be $\sim 10^{-3}$ per year in a galaxy.
Furthermore, no gamma ray bursts associated with the bursts
have been detected. 
To find progenitors of the bursts, several models\cite{model} have been proposed.

In the letter, we show that FRBs arise from collisions
between neutron stars and axion stars\cite{axion,axions}. 
The axion star is a boson star ( known as oscillaton\cite{axion2} ) made of axions bounded gravitationally.
The axion stars have been discussed\cite{kolb} to be formed in an epoch after the period of equal matter and
radiation energy density.

The production mechanism 
of FRBs is shown in the following. Under strong magnetic fields of neutron stars,
axion stars generate oscillating electric fields\cite{osc}. When they collide with the neutron
stars, the oscillating electric fields rapidly produce radiations 
in the atmospheres of the neutron stars.
Since the frequency of the oscillating electric field 
is given by the axion mass $m_a$,  
the frequency of the radiations produced\cite{iwa} 
in the collisions is equal to $m_a/2\pi\simeq 2.4\,\mbox{GHz}\,(m_a/10^{-5}\mbox{eV})$. 
This is the case of electrons accelerated by the
electric field initially stopping relative to the axion stars.
Actually, we need to take into account Doppler effect owing to the electron motions
in the atmosphere of the neutron stars, in order to explain finite band width of the observed FRBs.
 
The observations of FRBs constraint the parameters of the axion stars, that is,
the mass of the axion and the mass of the axion stars.
The observed frequency ( $\simeq 1.4$ GHz ) of FRBs 
gives the mass ( $\simeq 6\times 10^{-6}$eV ) of the axion,
while the observed rate of the bursts ( $\sim 10^{-3}$per year in a galaxy ) gives 
the mass ( $\sim 10^{-11}M_{\odot}$ ) of the axion stars under the assumption that
halo of galaxy is composed of the axion stars.
Then, with the use of the theoretical formula\cite{axions,osc} 
relating radius $R_a$ to mass $M_a$ of the axion stars,
we can find the radius $R_a\sim 160$km. 
Since the relative velocity $v_c$ at the time when the collisions occur
is estimated to be a several ten thousand km/s, we find that
the durations of FRBs are given by $R_a/v_c\sim $ a few milliseconds.

\vspace{0.1cm}
First we explain the classical solutions of the axion stars obtained 
in previous papers\cite{axions,osc,iwa,osci}.
The solutions are found by solving classical equations of axion field $a(\vec{x},t)$ coupled with gravity.
Assuming the axion potential such that $V_a=-f_a^2m_a^2\cos(a/f_a)\simeq -f_a^2m_a^2+m_a^2a^2/2$
for $a/f_a\ll 1$, 
we approximately obtain spherical symmetric solutions,

\begin{equation}
\label{a}
a(\vec{x},t)=a_0f_a\exp(-\frac{r}{R_a})\sin(m_at),
\end{equation} 
with $r=|\vec{x}|$,
where $m_a$ and $f_a$ denote the mass and decay constant of the axion, respectively.
The solutions represent boson stars made of the axions bounded gravitationally, named as axion stars.
The solutions are valid for the axion stars with small masses $M_a\ll 10^{-5}M_{\odot}$.
The radius $R_a$ of the axion stars is numerically given in terms of the mass $M_a$ by

\begin{equation}
\label{R}
R_a=6.4\frac{m^2_{\rm pl}}{m_a^2M_a}
\simeq 160\,\mbox{km}\,\,\Big(\frac{10^{-5}\rm eV}{m_a}\Big)^2\Big(\frac{10^{-11}M_{\odot}}{M_a}\Big), 
\end{equation} 
with the Planck mass $m_{\rm pl}$.  
The coefficient $a_0$ is given by 

\begin{equation}
\label{a_0}
a_0\simeq 2.2\times 10^{-6} \Big(\frac{10^2\,\mbox{km}}{R_a}\Big)^2\frac{10^{-5}\mbox{eV}}{m_a}.
\end{equation}
Thus, the condition $a/f_a\ll 1$ is satisfied for the axion stars with small mass $M_a \sim 10^{-11}M_{\odot}$.
Although we have used the mass $10^{-11}M_{\odot}$ for reference, the mass is the one we obtain
in the present paper.
We note that the maximum mass $M_{\rm max}$ of the stable axion stars is 
given\cite{osci} such that 
$M_{\rm max}\sim 0.6\,m_{\rm pl}^2/m_a\simeq 0.5\times 10^{-5}M_{\odot}(10^{-5}\mbox{eV}/m_a)$. 
Thus, the solutions represent stable axion stars.

The solutions as well as the parameters $R_a$ and $a_0$ can be easily found\cite{axions,osci}
for the axion stars with such small masses. In particular, the field configuration of the axion
can be interpreted as a wave function of the axion bounded by the gravitational force of the axion stars.
Then, 
the wave function $a(t,r)$ may take the form $a(t,r)=a_0f_a\sin(\omega t)\exp(-kr)$ with
the energy eigenvalue $\omega=\sqrt{m_a^2-k^2}\simeq m_a-k^2/2m_a$;
$k$ gives the radius of the axion stars, $k=R_a^{-1}\ll m_a$.
Obviously, the parameter $k^2/2m_a$ represents a binding energy of the axion which is roughly 
equal to $Gm_aM_a/R_a$
with the gravitational constant $G$. Thus, we obtain the radius $R_a\sim  m_{\rm pl}^2/2m_a^2M_a$ given in eq(\ref{R}).
Furthermore, the mass $M_a$ of the axion stars is approximately equal to 
the energy of the axion field 
$M_a=\int d^3x ((\vec{\partial}a)^2+m_a^2a^2)/2\simeq \int d^3x \,m_a^2a^2/2\simeq \pi m_a^2a_0^2f_a^2R_a^3/4$
where the average is taken in time. 	
Using the relation $m_a\simeq 6\times 10^{-6}\mbox{eV}\times (10^{12}\mbox{GeV}/f_a)$,
we can obtain the coefficient $a_0$ given in eq(\ref{a_0}).
In this way the solutions along with the parameters $R_a$ and $a_0$ can be approximately obtained. 
Hereafter, the precise values of these parameters are irrelevant for our discussions below.

\vspace{0.1cm}
Now, we proceed to discuss FRBs arising from the collisions between axion stars and neutron stars. 
The axion stars generate electric field $\vec{E}_a$ under magnetic field $\vec{B}$ of the neutron stars because
the axion couples with both electric field $\vec{E}$ and magnetic field $\vec{B}$ in the following,

\begin{equation}
\label{L}
L_{aEB}=k\alpha \frac{a(\vec{x},t)\vec{E}\cdot\vec{B}}{f_a\pi}
\end{equation}
with the fine structure constant $\alpha\simeq 1/137$,   
where the numerical constant $k$ depends on axion models; typically it is of the order of one.
Hereafter we set $k=1$.
From the Lagrangian, we derive the Gauss law, 
$\vec{\partial}\cdot \vec{E}=-\alpha\vec{\partial}(a\vec{B})/f_a\pi$. Thus,
the electric field generated by the axion stars under the magnetic field $\vec{B}$
is given by 

\begin{equation}
\label{ele}
\vec{E}_a(r,t)=-\alpha \frac{a(\vec{x},t)\vec{B}}{f_a\pi}=-\alpha \frac{a_0\exp(-r/R_a)\sin(m_at)\vec{B}}{\pi}.
\end{equation} 
The electric field oscillates coherently over the whole of the axion stars.
The field induces the oscillating electric currents with large length scale ( $\sim $ radius of neutron stars )
in the atmosphere of the neutron stars. Thus the large amount of radiations are emitted
so that the axion stars loses their energies.  
As will be shown later, the whole energies of a part of the axion stars 
touching the atmosphere of the neutron stars  
are released into the radiations.
Namely, the neutron stars make the axion stars evaporate into the radiations. 
The frequency of the radiations is given by $m_a/2\pi\simeq 2.4\times(10^{-5}\mbox{eV}/m_a)$GHz.
Since the collisions take place
over a short period $R_a/v_c$, we may identify the radiations as FRBs.

Before estimating how rapidly the axion stars lose their energies,
we calculate the rate of the collisions between axion stars and neutron stars in a galaxy.
Then, we can determine the mass of the axion stars by the comparison of theoretical
with observed rate of the bursts.
We assume that halo of a galaxy is composed of the axion stars whose
velocities $v$ relative to neutron stars is supposed to be $3\times 10^2 $\,km/s. 
Since the local density of the halo is supposed to be $0.5\times 10^{-24}\,\mbox{g\,cm}^{-3}$,
the number density $n_a$ of the axion stars is given by $n_a=0.5\times 10^{-24}\,\mbox{g\,cm}^{-3}/M_a$. 
The event rate $R_{\rm burst}$ can be obtained in the following,

\begin{equation}
R_{\rm burst}=n_a\times N_{\rm ns}\times Sv\times 1\rm year,
\end{equation}
where $N_{\rm ns}$ represents the number of neutron stars in a galaxy;
it is supposed to be $10^9$.
The cross section $S$ for the collision is given by
$S=\pi (R_a+R)^2\big(1+2G(1.4M_{\odot})/v^2(R_a+R)\big)$ 
where $R\,(=10$km ) denotes the radius of neutron star
with mass $1.4M_{\odot}$. 
It follows that the observed event rate is given by
$R_{\rm burst}\sim 10^{-3}\times(10^{-11}M_{\odot}/M_a)^2$ per year in a galaxy
when we take $M_a=10^{-11}M_{\odot}$. 
The radius of the axion stars with the mass is equal to $160$\,km, which is much larger than that of
the neutron star.
Thus, when the collisions take place, the neutron stars pass through the insides of the axion stars.
Then,
the amount of the radiation energy released in the collision is given by
$10^{-11}M_{\odot}(10\rm km/160\rm km)^2\sim 10^{43}$GeV.
It is coincident with
the observed energies of FRBs.
It should be noted that the mass $10^{-11}M_{\odot}$ of the axion stars is in the range
allowed as masses of axion miniclusters\cite{mini}, which eventually condense to axion stars. 

\vspace{0.2cm}
Now, we estimate how rapidly the axion stars emit radiations in the collisions
with neutron stars. In particular, 
we show that they rapidly lose their energies in the atmosphere of the neutron stars.
The electric field $\vec{E}_a$ in eq(\ref{ele}) 
generated under the magnetic field makes an electron with momentum $\vec{p}$ oscillate according to the
equation of motion $\dot{p}=eE_a$ where we write down 
the component of the equations parallel to the magnetic field $\vec{B}$.
Namely, the direction of the oscillation is parallel to $\vec{B}$.  
The emission rate of the radiation energy produced by the electron is
given by $2e^2\dot{p}^2/(3 m_e^2)$ with electron mass $m_e$, where we assumed dipole radiation.
When the number of electrons emitting the radiations is $N_e$,
the total emission rate is given by $2N_e e^2\dot{p}^2/(3 m_e^2)$.
However the oscillation of electrons in the axion star is coherent
so that the amount of radiation energy produced by the coherent oscillation is
given by $2e^2\dot{p}^2N_e^2/(3 m_e^2)$ since coherently oscillating electric current 
$epN_e/m_e$ is induced. 

We suppose radiations arising from a region with volume $(1\mbox{cm})^3$ in the atmosphere
of neutron stars. The size of the region is less than 
the wave length $(m_a/2\pi)^{-1}\simeq 10$ cm of the radiations.
Then, we obtain the emission rate $\dot{W}$ of the radiation energy,

\begin{equation}
\dot{W}=\frac{2(e\dot{p}N_e)^2}{3 m_e^2}
=\frac{2N_e^2}{3}\frac{e^2\alpha^2 a_0^2B^2}{\pi^2 m_e^2}
\sim 10^{-4}N_e^2\mbox{GeV/s}\,\Big(\frac{B}{10^{12}\rm G}\Big)^2 
\sim 10^{36}\,\mbox{GeV/s}\,\Big(\frac{n_e}{10^{20}\mbox{cm}^{-3}}\Big)^2\,\Big(\frac{B}{10^{12}\rm G}\Big)^2
\end{equation}
where the number of electron $N_e$ in the region is

\begin{equation}
N_e=(1\mbox{cm})^3 n_e
=10^{20}\, \Big(\frac{n_e}{10^{20}\mbox{cm}^{-3}}\Big),
\end{equation}
where we have taken as an example the number density $n_e=10^{20}/\rm cm^3$ of electrons realized in 
ionized hydrogen atmospheres with the density $\sim 10^{-4}$g/cm$^3$ of neutron stars.
On the other hand, the energy of the axion star contained in the region
is given by $10^{-11}M_{\odot}(1\mbox{cm})^3/(4\pi R_a^3/3)\sim 10^{24}\mbox{GeV}$.
The energy is smaller than the energy of the radiations
$\dot{W}\times (1\mbox{cm}/c\sim 10^{-10}\rm s)\simeq 10^{26}$GeV emitted within the time 
$1\mbox{cm}/c\sim 10^{-10}$s which it takes for the light to pass the distance $1$cm.
( $c$ denotes the light velocity. )
It implies that
the axion energy is immediately transformed into the radiation energy.
Therefore, we find that the axion stars lose their energies by emitting
the radiations near the surface of the neutron stars.
We note that the number density of electrons $n_e$ increases zero to 
much larger ones than $10^{20}/\rm cm^3$ as we go deeper inside
the atmosphere from the outside. 
It is notable that the total amount of the radiation energy 
$10^{-11}M_{\odot}(10\rm km/160\rm km)^2\sim 10^{43}$GeV produced in the collision
is coincident with the observed one.

It should be noticed that since the velocity $v_c$ with which the neutron star collides against the axion star 
is given by $v_c=\sqrt{2G(1.4M_{\odot})/R_a}\simeq 1.7\times 10^{-1}\simeq 5\times 10^4$km/s,
the time which it takes for the neutron star to pass the axion star is equal to 
$320\mbox{km}/v_c\simeq 6$ms. 
This gives the observed durations of FRBs.

We mention that
the magnetic fields of the neutron stars do not disappear in the collisions because the field energy 
$4\pi R^3B^2/3\sim 10^{44}\mbox{GeV}(B/10^{12}G)^2$ is larger than 
the axion energy $10^{-11}M_{\odot}(10\rm km/160\rm km)^2\sim 10^{43}$GeV
released in the collisions.

\vspace{0.1cm}
Therefore, comparing the theoretical formulas of the axion
stars with the observed values of FRBs, we conclude that
FRBs arise from the collisions between the axion stars and the neutron stars.  
It is remarkable that we can determine the mass of the axion by observing the frequencies of FRBs.
The axion mass $\simeq 6\times 10^{-6}$eV we obtain 
is in the window allowed by observational and cosmological
constraints\cite{axionc}.

\vspace{0.2cm}
The author
expresses thanks to Prof. J. Arafune for useful comments
and discussions.




\begin{thebibliography}{99}
\bibitem{frb}D. R. Lorimer, M. Bailes, M. A. McLaughlin, D. J. Narkevic, F. Crawford, 
Science, 318 (2007) 777.  \\
E. F. Keane, D. F. Ludovici, R. P. Eatough, et al.,
MNRAS, 401 (2010) 1057.
\bibitem{frb1}D. Thornton, B. Stappers, M. Bailes, et al. Science, 341 (2013) 53.
\bibitem{frb2}L. G. Spitler, J. M. Cordes, J. W. T. Hessels, et al., ApJ, 790 (2014) 101.
\bibitem{model}T. Totani, PASJ, L21 (2013) 65.\\
K. Kashiyama, K. Ioka, and P. M´esz´aros, ApJL, L39 (2013) 776.\\
S. B. Popov and K. A. Postnov arXiv:1307.4924.\\
H. Falcke and L. Rezzolla, A and A, A137 (2014) 562.\\
A. Loeb, Y. Shvartzvald and D. Maoz, MNRAS, L46 (2014) 439.\\
K. W. Bannister and G. J. Madsen, MNRAS, 353 (2014) 440.
\bibitem{axion}R. D. Peccei and H. R. Quinn, Phys. Rev. Lett. 38 (1977) 1440.\\
S. Weinberg, Phys. Rev. Lett. 40 (1978) 223.\\
F. Wilczek, Phys. Rev. Lett. 40 (1978) 279.
\bibitem{axions}P. Jetzer, Phys. Rep. 220 (1992).\\
E. Seidel and W.M. Suen, Phys. Rev. Lett. 66 (1991) 1659.\\
A. Iwazaki, Phys. Lett. B451 (1999) 123.
\bibitem{axion2}E. Seidel and W.M. Suen, Phys. Rev. Lett. 72 (1994) 2516.
\bibitem{kolb}E. W. Kolb and I. I. Tkachev, Phys. Rev. Lett. 71 (1993) 3051.
\bibitem{osc}A. Iwazaki, Prog. Theor. Phys. 101 (1999) 1253.\\
A. Iwazaki, Phys. Rev. D60 (1999) 025001.
\bibitem{iwa}A. Iwazaki, hep-ph/9908468;
Phys. Lett. B455 (1999) 192;\\
Phys. Lett. B486 (2000) 147;
Phys. Lett. B489 (2000) 353.
\bibitem{osci}M. Alcubierre, R. Becerril, F. S. Guzman, T. Matos, D. Nu˜nez and L. A. Ure˜na-L´opez,
Class.Quant.Grav. 20 (2003) 2883.
\bibitem{mini}E. W. Kolb and I. I. Tkachev, Phys. Rev. Lett. 71 (1993) 3051;
Astrophys. J. 460 (1996) L25.
\bibitem{axionc}M. S. Turner, Phys. Rept. 197 (1990) 67.\\
J. E. Kim, and G. Carosi, Rev. Mod. Phys. 82 (2010) 557.
\end{thebibliography}
\end{document}